\documentstyle[12pt,epsfig,amsmath,axodraw]{article}
\def\lsim{\raise0.3ex\hbox{$\;<$\kern-0.75em\raise-1.1ex\hbox{$\sim\;$}}}
\def\gsim{\raise0.3ex\hbox{$\;>$\kern-0.75em\raise-1.1ex\hbox{$\sim\;$}}}
\def\neut{\chi_1^0}

\def\tb{\tan\beta}
\def\s{\smallskip}

\begin{document}

\vspace{-1truecm}

\rightline{LPT--Orsay 06/46}
\rightline{hep-ph/0609234}
\rightline{September 2006}   

\vspace{1.cm}

\begin{center}

{\Large {\bf The Higgs intense--coupling regime in constrained }}

\vspace{0.3cm}

{\Large {\bf SUSY models and its astrophysical implications}}

\vspace{1.cm}

{\large {\sc Abdelhak Djouadi} and {\sc Yann Mambrini} }

\vspace{1.cm}

Laboratoire de Physique Th\'eorique, CNRS and Universit\'e Paris--Sud, \\
Bt. 210, F--91405 Orsay Cedex, France.

\vspace{0.2cm}

\end{center}

\vspace{0.5cm}

\begin{abstract} We analyze the Higgs intense--coupling regime, in which all
Higgs particles of the Minimal Supersymmetric Standard Model are light with
masses of the same order and the value of $\tb$ the ratio of vacuum expectation
values of the two Higgs fields is large, in the framework of Supergravity
scenarios with non--universal soft Supersymmetry breaking scalar masses in
the Higgs sector.  In particular, we calculate the relic density abundance of
the lightest neutralino candidate for cold dark matter and  the rates in direct
and indirect detection at present and future experiments. We first show that
while in the mSUGRA model this regime is disfavored by present data, there are
regions in the parameter space of models with non--universal Higgs masses where
it can occur.  We then show that because of the large value of $\tan\beta$ and
the relatively low values of the neutral Higgs boson masses, the cross section
for neutralino--nucleon scattering is strongly enhanced in this regime and
would allow for the observation of a signal in direct detection experiments
such as CDMS--Soudan.  The expected sensitivity of gamma--ray detectors like
GLAST might be also sufficient to observe the annihilation of neutralinos in
such a regime.  \end{abstract}

\newpage

\section{Introduction}

In the Minimal Supersymmetric Standard Model (MSSM) \cite{SUSY}, the scalar
sector is extended to include two Higgs doublets fields to achieve the breaking
of the ${\rm SU(2)_L \times U(1)_Y}$ electroweak gauge symmetry. This leads to
the existence of five Higgs particles:  two CP--even Higgs bosons $h$ and $H$,
a CP--odd or pseudoscalar Higgs boson $A$, and two charged Higgs particles
$H^\pm$ \cite{SUSY,Review}. At the tree--level, the spectrum is determined by
two basic parameters: the mass of the pseudoscalar Higgs boson $M_A$ and the
ratio of the vacuum expectation values of the Higgs fields, $\tb =v_2/v_1$.  If
the pseudoscalar Higgs particle is very heavy, $M_A \gg M_Z$, one is in the
so--called decoupling regime \cite{decoupling} in which the lightest CP--even
$h$ particle is SM--like and has a mass that is close to the $Z$ boson mass
while the other CP--even Higgs particle $H$ and the charged $H^\pm$ bosons are
very heavy and degenerate in mass with the pseudoscalar $A$ boson, $M_H \approx
M_{H^\pm} \approx M_A$. The MSSM Higgs sector reduces then to the one of the
SM, but with a Higgs particle  that it rather light: $M_h^{\rm max} \approx
M_Z$ at the tree--level but $M_h^{\rm max} \sim 100$--140 GeV, depending on
$\tb$ and the strength of the important radiative corrections which need to be
included \cite{Review,RC}.  In the opposite scenario, in which $M_A \lsim M_Z$
\cite{antidecoupling} called the anti--decoupling regime in Ref.~\cite{Review},
it is the lighter CP--even $h$ which is degenerate in mass with the $A$ boson,
while the heavier $H$ boson is SM--like with a mass close to  $M_H^{\rm min}
\simeq M_h^{\rm max}$.  \s

An intermediate scenario, the so--called intense--coupling regime
\cite{intense}, is characterized by a rather large value of $\tb$ and a mass
for the pseudoscalar $A$ boson that is close but not equal to the maximal
(minimal) value  of the CP--even $h$ ($H$) boson mass.  In such a scenario, an
almost mass degeneracy of the neutral Higgs particles of the model occurs, $M_h
\approx M_A \approx M_H \approx 100$--140 GeV, while the couplings of both the
CP--even $h$ and $H$ particles to gauge bosons and isospin up--type fermions
are suppressed, and their couplings to down--type fermions, and in particular
$b$--quarks and $\tau$ leptons, are strongly enhanced. The interactions of both
Higgs particles therefore approach those of the pseudoscalar Higgs boson which
does not couple to massive gauge bosons as a result of CP invariance, and for
which the couplings to isospin $-\frac12 \, (+\frac12)$ fermions are
(inversely) proportional to $\tb$.\s

This scenario leads to a very interesting collider phenomenology which has been
discussed in detail in Ref.~\cite{intense}. In particular,  it has been shown
that one has to face a rather difficult situation for the detection of these
particles at the LHC, since the branching ratios of the usual interesting
decays which allow the detection of the CP--even Higgs bosons are too small and
cannot be used anymore, while the dominant $b\bar b$ and $\tau^+ \tau^-$ decay
modes have too large backgrounds. Even when using some rare Higgs decays, the
detection of the three individual Higgs bosons is very challenging in general
and in some cases even impossible at the LHC. At the future International
Linear $e^+e^-$ Collider (ILC), the three peaks can be resolved but the
measurement of the masses of the particles is more difficult than in other
scenarios.\s 

The studies on the intense coupling regime mentioned above have been performed
only in the framework of an effective low--energy MSSM in which the parameters
which break softly Supersymmetry (SUSY) are incorporated by hand, leading to a
Higgs sector that is practically disconnected from the SUSY particle sector. 
However, it would be more interesting to consider constrained SUSY models,
which are theoretically more appealing and which have a smaller number of basic
input parameters. A scenario that is widely used as a benchmark for constrained
Grand Unified SUSY Theories (SUSY GUTs), is the minimal Supergravity model
(mSUGRA) \cite{msugra} in which there are only five input parameters: a
universal value $m_0$ for the soft SUSY--breaking masses of all the scalars, a
universal soft SUSY--breaking gaugino mass parameter $M_{1/2}$, a common
trilinear Higgs--sfermion coupling $A_0$ [the three of which are defined at 
the GUT scale, $M_{\rm GUT} \sim 2 \times 10^{16}$ GeV], $\tb$ and the sign of 
the Higgs--higgsino mass parameter $\mu$.  An interesting question is whether 
the intense coupling regime can occur in such a scenario.\s 

A second question that one might ask is whether this scenario is compatible
with cosmological observations and, in particular, with the observed amount of
dark matter in the universe. Indeed, it is now well established that luminous
matter makes up only a small fraction of the observed mass in the universe and
a weakly interacting massive particle, identified in the MSSM with the stable
lightest neutralino, is one of the leading candidates for the ``dark''
component which is needed to explain astrophysical data on the rotation curves
of large scale structures of the universe \cite{reviewDM} and which has been
recently measured via the anisotropies of the cosmic microwave background by
the WMAP satellite with a very good precision \cite{wmap03-1}. A related
question is whether in such a scenario, the lightest neutralino of the MSSM
would lead to observable effects in the present and near future experiments
which are devoted to the direct and indirect detection of these dark matter
particles.\s 

In this paper, we attempt to answer to these questions. We show that in
the framework of the mSUGRA model, the particle spectrum is so constrained 
that the intense coupling regime is realized only in a very limited area of the
parameter space; in addition, because it occurs only for extremely large values
of $\tb$, the obtained spectrum could potentially lead to a conflict with
experimental data from collider physics. We then investigate the Non Universal
Higgs Mass (NUHM) model,  discussed in Refs.~\cite{NUHM1,NUHM2} for instance, in
which the equality of the soft SUSY--breaking masses of the two Higgs fields 
$m_{H_1}$ and $m_{H_2}$ with the common mass term $m_0$ for squarks and sleptons
at the GUT scale is relaxed.  Such a non--universality structure might, for 
instance, occur in SUSY GUTs in which the fields $H_1$ and $H_2$ belong to
different multiplets \cite{NUHM1};  it can also be obtained in the low--energy 
limit of some phenomenologically appealing string scenarios such as heterotic 
orbifold models \cite{Brignole} and effective heterotic  Anomaly Mediated SUSY 
Breaking scenarios \cite{Pierre}. The NUHM model is then equivalent to an 
effective low energy MSSM where, in addition to the mSUGRA parameters, on has 
the higgsino mass parameter $\mu$ and the pseudoscalar Higgs boson mass $M_{A}$
as free input parameters. This additional freedom will allow to realize the 
intense Higgs coupling regime, while complying with all known accelerator and 
astrophysical data, in a much larger region of the parameter space compared 
to the mSUGRA case.\s

The paper is organized as follows. In section 2 we review the NUHM model
and discuss the impact of non--universal scalar mass parameters on the Higgs 
sector as well as on the cosmological relic density of the dark matter lightest
neutralino and on its detection in direct and indirect searches. In section 3, 
we analyze the Higgs intense regime in the NUHM model: we first explain the 
procedure to obtain the spectrum and the various constraints that we impose on 
it, and then present two concrete examples in which this regime is realized; 
we then discuss the implications of these scenarios for the relic density of 
the neutralino and its direct and indirect detection rates. For completeness, 
we briefly discuss in section 4, the case of the mSUGRA model. A brief 
conclusion  is given in section 5.

\subsection*{2. The NUHM model}

\subsubsection*{2.1 The impact of non--universality on the Higgs sector}

Let us first recall that in the mSUGRA model, in addition to the four basic
continuous input parameters, the common soft SUSY--breaking scalar $m_0$ and
gaugino $M_{1/2}$ mass parameters and trilinear coupling $A_0$, defined at the
GUT scale, and the ratio of vevs $\tan \beta$,  the sign of the Higgs--higgsino
parameter $\mu$ is also free. This is due to the fact that $\mu^2$ is
determined by the minimization of the scalar Higgs potential which leads to
electroweak symmetry breaking (EWSB). At the tree level, one has in terms of 
$\tb$ and the two soft SUSY--breaking Higgs mass terms defined at the SUSY 
breaking scale $M_S = {\cal O}(1$ TeV), 
\begin{equation}
  \mu^2 = \frac{m^2_{H_1} - m^2_{H_2} \tan^2 \beta}{\tan^2 \beta -1 } - 
  \frac{1}{2} M_Z^2\ .
  \label{electroweak}
\end{equation} 
\noindent which, for reasonably large values of  $\tan\beta$ ($\gsim 5$),  can
be  approximated by
\begin{equation}
\mu^2\approx -m_{H_2}^2-\frac{1}{2} M_Z^2\ ,
\label{electroweak2}
\end{equation}
with $m_{H_2}^2$ being negative to effectively generate EWSB. Again at the 
tree level, the mass of the CP--odd Higgs boson is approximately given by
\begin{equation}
M^2_A=m_{H_1}^2+m_{H_2}^2+2\mu^2\ ,
\label{ma} 
\end{equation}
which can be rewritten using eq.~(\ref{electroweak2})
\begin{equation}
M^2_A \approx m_{H_1}^2-m_{H_2}^2-M_Z^2\ ,
  \label{ma2} 
\end{equation}
The masses of the neutral CP--even Higgs bosons and the charged Higgs particle
are then obtained from the usual tree--level relations 
\begin{eqnarray}
M_{h,H}^2 &=& \frac{1}{2} \left[ M_A^2+M_Z^2 \mp \sqrt{ (M_A^2+M_Z^2)^2 -4M_A^2
M_Z^2 \cos^2 2\beta } \right] \nonumber \\
M_{H^\pm}^2&=& M_A^2 + M_W^2 
\label{Hmasses:tree}
\end{eqnarray}
Of course, to have a more accurate determination of all the Higgs boson masses
and proper EWSB, it is important that the radiative corrections are included; 
see Refs.~\cite{Review,RC} for reviews.\s

In the non--universal Higgs mass (NUHM) model, the universality of  the soft
SUSY--breaking Higgs mass parameters at the GUT scale is relaxed, that is 
$m_{H_1} \neq m_{H_2} \neq m_0$.   This non--universality might, for instance, 
occur  in GUT constructions where the fields $H_1$ and $H_2$ belong to different
multiplets. This model is then equivalent to the ``effective low energy" MSSM 
where in addition to the mSUGRA parameters, on has $\mu$ and $M_{A}$ as free
input parameters. In this paper, we will parameterize  this non--universality  
through two dimensionless parameters $\delta_1$ and $\delta_2$, which measure 
the relative  deviation from the mSUGRA case at the GUT scale, as follows
\begin{equation} 
m_{H_{1}}^2(M_{\rm GUT})=m_0^{2} (M_{\rm GUT}) (1+\delta_{1})\ , \quad 
m_{H_{2}}^2(M_{\rm GUT})=m_0^{2} (M_{\rm GUT}) (1+ \delta_{2})\  
\label{Higgsespara} 
\end{equation} 
from which one can clearly see that lower (larger) values of $m_{H_1}$
($m_{H_2}$), corresponding to $\delta_1 < 0$ ($\delta_2 > 0)$, imply  a lighter
$A$ boson. Thus, contrary to the mSUGRA case where the CP--odd $A$ boson is very
heavy for reasonable values of $m_0$ [except in the so--called focus point
region where $m_0$ and $\tan \beta$ are very large, see Ref.~\cite{Feng} for
instance], here, one can  use the additional freedom and  adjust the parameters
$\delta_{1,2}$ in such way that the obtained value of $M_A$ is not too large. In
particular, for high values of $\tan\beta$, one can realize the scenario in
which the Higgs sector of the  model is in the intense coupling regime, i.e.
$M_A \approx 100$--130 GeV which  leads to the relation $M_A \approx M_H \approx
M_h$.\s 

Such a departure from universality leads to drastic consequences not only in 
the Higgs sector, but also in the chargino and neutralino sectors as the Higgs 
mass parameters $m_{H_1}$ and $m_{H_2}$ enter in the determination of $\mu$ 
which enters as a basic input parameter in the mass matrices of these 
states. This would for instance alter the nature of the dark matter lightest
neutralino  with a possibly large impact on its cosmological relic density 
abundance  and its detection rates, as will be summarized in the next 
subsections.

\subsubsection*{2.2 Consequences on the LSP relic density}

In the MSSM there are four neutralinos, ${\chi}^0_i$ with $i=1,2,3,4$, among  
which the lightest one ${\chi}^0_1$ is the lightest SUSY particle (LSP) and 
the SUSY candidate  for the cold dark matter in the universe. Being a 
superposition of the bino, wino and higgsino fields, respectively denoted by 
$\widetilde{B}^0, \widetilde{W}_3^0$ and $\widetilde{H}^0_1,\widetilde{H}_2^0$,
the $\neut$ neutralino texture can be written as
\begin{equation}
\chi^0_1 = {Z_{11}} \widetilde{B}^0 + {Z_{12}} \widetilde{W}_3^0 +
{Z_{13}} \widetilde{H}^0_2 + {Z_{14}} \widetilde{H}^0_1\ .
\label{lneu}
\end{equation}
where $Z_{ij}$ are elements of the matrix which diagonalizes the $4\times 4$ 
neutralino mass matrix. It is commonly defined that ${\chi}^0_1$ is mostly 
gaugino--like  if $P\equiv \vert {Z_{11}} \vert^2 + \vert {Z_{12}}  \vert^2 > 
0.9$,  higgsino--like if $P<0.1$, and a mixed state otherwise. In the NUHM 
scenario, one can choose the values of $m_{H_1}$ and $m_{H_2}$ at the GUT scale
in such a way that the electroweak symmetry breaking conditions lead to a low 
value for $\mu$. As a consequence, the lightest neutralino is generally 
higgsino--like or a mixed bino--higgsino state.\s

This has very important consequences on the relic density, which is inversely
proportional to the LSP annihilation cross section. In Fig.~1, we display the
Feynman  diagrams of the main processes contributing to the annihilation cross
section of the neutralinos to fermions (a), $\neut \neut \to f\bar f$, and to 
gauge or Higgs bosons (b), $\neut \neut \to VV$ and $\Phi_i\Phi_j$ with $V=W,
Z$ and $\Phi_i=h,H,A$ and $H^\pm$, together  with the  expressions of the
relevant parts of the amplitudes \cite{DM-DN}. From these diagrams and the 
LSP neutralino texture given by eq.~(\ref{lneu}), one can make the following  
remarks.\s 

\begin{figure}[!t]
\vspace*{-.5cm}
\begin{center}
\begin{picture}(100,90)(-30,-5)
\hspace*{-11.5cm}
\SetWidth{1.1}
\Text(120,75)[]{\bf a)}
\Line(150,25)(250,25)
\Line(150,75)(250,75)
\DashLine(200,25)(200,75){4}
\Text(142,30)[]{$\chi_1^0$}
\Text(142,70)[]{$\chi_1^0$}
\Text(207,50)[]{$\tilde f$}
\Text(255,35)[]{$f$}
\Text(255,70)[]{$\bar f$}
\hspace*{4.8cm}
\ArrowLine(150,25)(185,50)
\ArrowLine(150,75)(185,50)
\DashLine(185,50)(230,50){4}
\ArrowLine(230,50)(265,25)
\ArrowLine(230,50)(265,75)
\Text(142,30)[]{$\chi_1^0$}
\Text(142,70)[]{$\chi_1^0$}
\Text(210,65)[]{$h,H,A$}
\Text(270,40)[]{$f$}
\Text(270,65)[]{$\bar f$}
\hspace*{5.2cm}
\ArrowLine(150,25)(185,50)
\ArrowLine(150,75)(185,50)
\Photon(185,50)(230,50){3.5}{5}
\ArrowLine(230,50)(265,25)
\ArrowLine(230,50)(265,75)
\Text(142,30)[]{$\chi_1^0$}
\Text(142,70)[]{$\chi_1^0$}
\Text(210,65)[]{$Z^*$}
\Text(270,40)[]{$f$}
\Text(270,65)[]{$\bar f$}
\Text(100,0)[]{$\propto \frac{m_{\chi} m_f}{m_{\tilde f}^2}Z_{11}^2~~~$ 
\hspace{1cm}
$\propto \frac{m_{\chi}^2}{M_A^2} \frac{Z_{11}Z_{13,14}}{m_W}  
m_{f} (\tan \beta)^{-2I_3^f} ~~~~~~$ \hspace{1cm}
$\propto \frac{m_f m_{\chi}}{m_Z^2} Z_{13,14}^2~$  \hspace{1cm}}
\end{picture}
\vspace*{.6cm}
\end{center}
\vspace*{-3mm}
\end{figure}
\setcounter{figure}{0}
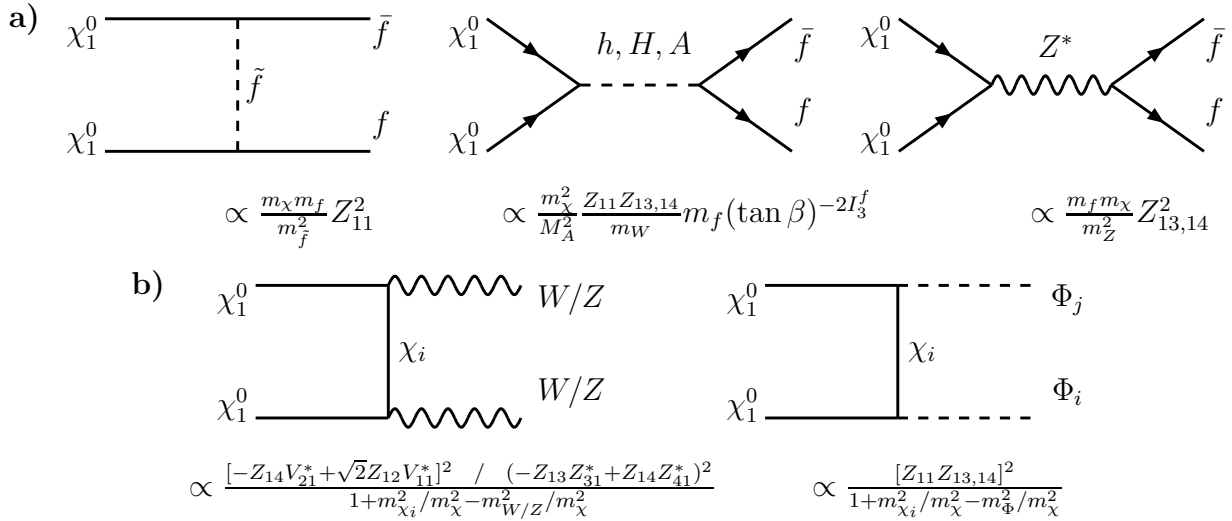
\begin{figure}[!h]
\vspace*{-1.2cm}
\begin{center}
\begin{picture}(100,90)(-30,-5)
\hspace*{-9.5cm}
\SetWidth{1.1}
\Text(110,75)[]{\bf b)}
\Line(150,25)(200,25)
\Line(150,75)(200,75)
\Line(200,25)(200,75)
\Photon(200,25)(250,25){3.5}{5}
\Photon(200,75)(250,75){3.5}{5.}
\Text(142,30)[]{$\chi_1^0$}
\Text(142,70)[]{$\chi_1^0$}
\Text(210,50)[]{$\chi_i$}
\Text(270,35)[]{$W/Z$}
\Text(270,70)[]{$W/Z$}
\hspace*{6.5cm}
\Line(150,25)(200,25)
\Line(150,75)(200,75)
\Line(200,25)(200,75)
\DashLine(200,25)(250,25){4}
\DashLine(200,75)(250,75){4}
\Text(142,30)[]{$\chi_1^0$}
\Text(142,70)[]{$\chi_1^0$}
\Text(210,50)[]{$\chi_i$}
\Text(265,35)[]{$\Phi_i$}
\Text(265,70)[]{$\Phi_j$}
\Text(100,-2)[]{$\propto \frac{[-Z_{14} V_{21}^* + \sqrt{2} Z_{12} V_{11}^*]^2 
~~/ ~~(-Z_{13} Z_{31}^* +  
Z_{14} Z_{41}^*)^2}{1+m_{\chi_i}^2 /
m_{\chi}^2 - m_{W/Z}^2 / m_{\chi}^2}$ \hspace*{1cm}
$\propto \frac{[Z_{11} Z_{13,14} ]^2}{1+m_{\chi_i}^2 /
m_{\chi}^2 - m_{\Phi}^2 / m_{\chi}^2}$
}
\end{picture}
\vspace*{-.2cm}
\end{center}
\caption[]{Feynman diagrams for LSP neutralino annihilation into a fermion pair
(a) and into massive gauge bosons and Higgs bosons (b). The relevant parts of 
the amplitudes are shown explicitly. $V$ and $Z$ are the chargino and 
neutralino mixing matrices.} 
\vspace*{-2.mm}
\end{figure}

\begin{itemize}

\item For LSP annihilation into light fermions [in general $\tau$--leptons and
to a lesser extent $b$ quarks], the diagram with $t$--channel sfermion exchange
contributes only if ${\chi}_1^0$ is bino--like [as higgsinos couple to fermions
proportionally to their masses], while for a higgsino--like LSP, the coupling
to the $Z$ boson is large [being proportional to $Z_{13}$ and $Z_{14}$],
enhancing thus the  annihilation channel $\neut \neut \xrightarrow{Z} f
\bar{f}$.  
\vspace*{-2mm}

\item The annihilation through Higgs boson exchange, $\neut \neut
\xrightarrow{A,h,H} f  \bar{f}$, contributes substantially only close to
the pole of the Higgs boson exchanged in the $s$--channel and if the LSP is a
mixed state in which case the coupling to the Higgs boson is large [both
the $Z_{11}$ and  $Z_{13}(Z_{14})$ matrix elements are large]; the dominant
final state is $b\bar b$ for which the amplitude, $\propto m_b \tan\beta$,  is
enhanced for large values of $\tan \beta$. Note that the dominant  component is
due to the exchange of the pseudoscalar $A$ boson as its $s$--wave contribution
is not suppressed by the small velocity of the neutralinos, in contrast to the
$p$--wave exchange of the CP--even $h$ and $H$ bosons. 
\vspace*{-2mm}

\item The annihilation into $WW/ZZ$ bosons is efficient only for  higgsino--like
or mixed neutralinos when the elements $Z_{13}(Z_{14})$ are large. There is also
an annihilation diagram through $s$--channel CP--even $h,H$ bosons exchange but
which gives only a small contribution as one is far from the Higgs 
boson poles and/or the Higgs couplings to neutralino and gauge bosons are 
suppressed; in addition, the $p$--wave contributions are suppressed by the
small velocity of the LSPs. 
\vspace*{-2mm}

\item Since in the intense coupling regime all the Higgs particles are light,
annihilation into a Higgs boson and a massive gauge boson, e.g. $\neut \neut
\to AZ$ or $H^\pm W^\mp$, and two Higgs bosons, e.g. $\neut \neut \to Ah$ or
$H^+ H^-$, can occur if the neutralino LSP is heavy enough, $m_{\chi^0_1} \gsim
150$ GeV. The dominant contribution is due to neutralino or chargino exchange;
the $Z (h,H)$ exchange diagrams give small contributions for $Ah (AZ)$
annihilation as one is far from $s$--channel poles  in this case. 
\vspace*{-2mm}

\item For higgsino--like LSPs,  the co--annihilation processes with the lightest
chargino and the next--to--lightest neutralino are very efficient because of 
the degeneracy $m_{\chi^{\pm}_1} \sim m_{\chi^0_2} \sim m_{\chi^0_1}$ and lead 
to the right density only for very heavy LSPs \cite{DM-CO1}. There are 
additional  diagrams which might contribute to the LSP cross section such as 
co--annihilation with the lightest sfermion \cite{DM-CO2} [in general 
$\tilde{\tau}_1$] when it is almost degenerate with the LSP. 
\end{itemize}

\subsubsection*{2.3 The impact on the detection of neutralinos}

Let us now discuss the expected signal of the intense coupling regime of the
NUHM model in the two most promising search strategies for weakly interacting
massive particle (WIMP) dark matter candidates:  the search of the elastic
scattering of ambient neutralinos off a nucleus in a laboratory detector through
nuclear recoils, the ``direct search''  \cite{Nonunivdirect,Complementarity}, 
and the search for interesting decays products of WIMP  annihilation into
standard particles such as gamma rays, positrons, neutrinos etc..,  the
so--called ``indirect detection'' \cite{Complementarity,Nonunivgamma}. 
\smallskip

The strength of the direct detection signal is directly proportional to the
neutralino--nucleon scattering cross section, $\sigma ({ \chi}_1^0 N \to {
\chi}_1^0 N)$.    The matrix element for this scattering, mediated by squark
and $Z$ boson exchange as well as Higgs boson exchange diagrams [the crossed
diagrams of those shown in Fig.~1a] receives both spin--dependent and
spin--indepen\-dent contributions. The former play a sub-dominant
role in most direct search experiments, which employ fairly heavy nuclei. The
spin--independent contribution in turn is usually dominated by Higgs boson
exchange diagrams, where the Higgs bosons couple either directly to light
($u,d,s$) quarks in the nucleon, or couple to two gluons through a loop of
heavy ($c,b,t$) quarks or squarks. Only scalar Higgs couplings to neutralinos
contribute in the non--relativistic limit and therefore,  in the absence of
CP--violation in the Higgs sector of the NUHM model discussed here, one only
needs to include the contributions of the two neutral CP--even $h$ and $H$
particles. The contribution of the CP--even Higgs boson which has enhanced
couplings to down--type quarks for $\tan\beta \gg 1$ is by far dominating. 
Thus, while in mSUGRA models only the contribution of the generally heavy $H$
boson is relevant, in the intense coupling regime, both the $h$ and $H$
particles will play a role. \smallskip

Many direct detection experiments have been and are presently carried out
around the world. Recent collaborations such as CDMS \cite{experimento2} and
EDELWEISS \cite{edelweiss} have explored the regions of parameter space
corresponding to a WIMP--nucleon cross section of $\sigma \gsim 10^{-6}$ pb. 
Future experiments will have improved sensitivity and, for example, GEDEON
\cite{IGEX3} will be able to explore  a WIMP--nucleon cross section of $\sigma
\gsim 3\times 10^{-8}$ pb. Similar cross sections will also be tested by the
EDELWEISS II experiment while CDMS--Soudan \cite{soudanmine} [an extension of
the CDMS experiment in the Soudan mine], will be able to test $\sigma \gsim
2\times 10^{-8}$ pb [in our  study, we will take the more conservative bound of
$4 \times 10^{-7}$ pb which corresponds to the present sensitivity of the
experiment].\smallskip

On the other hand, there are also promising methods for the indirect detection
of WIMPs through the analysis of their annihilation into SM particles [with 
some of the contributing diagrams are the same as those depicted in Fig.~1] and
the search  for their stable decay and fragmentation products,  i.e. neutrinos,
photons, protons, antiprotons, electrons and positrons \cite{Feng}.  While
electrons and protons are undetectable in the sea of matter particles in the
universe, neutrinos, photons, positrons and anti-protons could be detected over
the background due to ordinary particle interactions. An interesting possibility
consists of  detecting the gamma rays produced by these annihilations in the
galactic halo. For this purpose, one uses atmospheric Cherenkov telescopes or
space--based gamma--ray detectors.  Planed experiments will reach significant
sensitivity and, for instance, the GLAST telescope \cite{GLAST},  which is 
scheduled for launch in 2007, will be able to detect a flux of gamma rays from 
dark matter particles of the order of $\Phi_{\gamma}\sim 10^{-10}$ photons 
cm$^{-2}$s$^{-1}$.\smallskip

The detection of the neutralinos have been discussed at length in  the mSUGRA
scenario, see Ref.~\cite{reviewDM} for reviews. The rates  can be increased in
different ways when the structure of mSUGRA for the soft SUSY--breaking scalar
terms is abandoned in the Higgs sector
\cite{Nonunivdirect,Complementarity,Nonunivgamma}. In direct detection for
instance,  the neutralino--nucleon elastic cross section, where the exchange of
the CP--even Higgs bosons gives the dominant contribution as discussed above,
can be considerably enhanced when decreasing the pseudoscalar Higgs mass and,
thus, the heavy Higgs boson mass, through a particular choice of the
non--universality parameters $\delta_1$ and $\delta_2$. Concerning the
prospects for the indirect detection of the neutralinos, we have seen that it
is possible to enhance the annihilation channel involving the exchange of the
CP--odd Higgs boson $A$ by reducing its mass ($\delta_1 < 0, \delta_2 >0$) and
have a larger yield for the interesting decay products of the $b$--quarks in
the reaction $\neut \neut \xrightarrow{A} b \overline{b}$. In addition,  by
increasing the higgsino component of the lightest neutralino ($\delta_2 > 0$),
one would  have a larger cross section for the annihilation processes $\neut
\neut \rightarrow WW/H^+H^-$ for instance which allow to look for interesting
decay products of the gauge and Higgs bosons such as photons.  

\subsection*{3. The analysis}

\subsubsection*{3.1 The determination of the spectrum and constraints}

In the present analysis, we use the Fortran code {\tt SuSpect} \cite{Suspect}
to solve the (two--loop)  Renormalization Group Equations for the gauge and
Yukawa couplings and the soft SUSY--breaking parameters, to achieve proper
electroweak symmetry  breaking with the (two--loop) scalar potential,  and to
calculate the spectrum of the physical SUSY particles and Higgs bosons
including radiative corrections. We follow the procedure outlined in
Ref.~\cite{ddk}, except that we allow for non--universality for the soft
SUSY--breaking Higgs mass parameters as explained in the previous section.  In
addition to leading to a consistent electroweak symmetry breaking [i.e. 
obtaining a  reasonable value for $\mu, M_A$ and $M_Z$ as well as no charge or
color  breaking (CCB) minima] and to a  phenomenologically viable spectrum [no
tachyonic sparticles and Higgs bosons and and LSP that is not charged], a given
set of input parameters has to satisfy experimental constraints. The ones
relevant for this study are as follows \cite{PDG}.  

\begin{itemize}

\vspace*{-2mm}
\item[--]  The total cross section for the production of any pair of sparticles
  at the highest LEP energy (209 GeV) must be less than 20 fb. This leads  for
  instance to an upper bound of $\approx 100$ GeV on the chargino and slepton/
  3d generation squark masses.  The masses of the gluino and first/second
  generation squarks should be larger than $\approx 200$ GeV to cope with the
  Tevatron exclusion bounds. 

\vspace*{-2mm}  
\item[--]  Searches for neutral Higgs bosons at LEP  impose a lower bound on
  $M_h$ which, in the decoupling limit when $M_A \gg M_Z$, is close to 114 GeV. 
  Allowing for a theoretical uncertainty of $\approx 3$ GeV, one  requires 
  $M_h>$  111 GeV in this case. For small $M_A$ values, the bound should be
  of the order of $M_h \sim M_A \gsim M_Z$. 

\vspace*{-2mm}
\item[--] Quantum corrections from superparticles to electroweak observables 
can be incorporated through the $\rho$ parameter which should obey the upper 
bound $\delta \rho_{\rm SUSY} < 2.2 \cdot 10^{-3}$ at the $2\sigma$  level. 
However, it turns out that this constraint is always superseded by either the 
LEP Higgs search limit or by the CCB constraint.

\vspace*{-2mm}
\item[--]  Recent measurements  of the muon magnetic moment  lead to a
constraint on the SUSY contribution $ -5.7 \cdot 10^{-10} \leq a_{\mu}^{\rm
SUSY} \leq 4.7 \cdot 10^{-9}$ where, for the determination of the hadronic SM 
contributions, both data from $e^+e^-$ annihilation into hadronic final  states
and data from semileptonic $\tau$ decays are used, leading to less than
$1\sigma$ deviation between the experimentally measured  value and the 
theoretical prediction in the SM. 

\vspace*{-2mm} 
\item[--]   Allowing for experimental and theoretical errors,
the branching ratio for radiative $b$ decays should be $ 2.33 \cdot 10^{-4}
\leq {\rm B} (b \rightarrow s \gamma) \leq 4.15 \cdot 10^{-4}$. The branching
ratio for the very rare decay into $\mu^+ \mu^-$ should be bounded by B($B_s
\to \mu^+ \mu^-$) $< 2.9\times 10^{-7}$ \cite{bmumuexp} [this observable does
not yet constrain the parameter space of mSUGRA, but it it has been stressed
recently \cite{Nonunivdirect} that it should be  taken into account in the
non-universal case]. We note however, that both the $b \rightarrow s \gamma$
and  $B_s \to \mu^+ \mu^-$ constraints have a different status from those
discussed earlier, as a small amount of squark flavor mixing would make these
observable compatible with the experimental values while having a  negligible
effect  on the other (flavor conserving) observables, such as signals at
colliders.  

\vspace*{-2mm}
\end{itemize}

The theoretical constraints are implemented directly in the Fortran code {\tt 
SuSpect}; the program implements also the experimental constraints mentioned
above except for the rate B($B_s \to \mu^+ \mu^-$) which we determine using the
code {\tt microMEGAS} \cite{micromegas}. \bigskip

In addition to these constraints from ``collider experiments", we also require
that the calculated $ \chi_1^0$ cosmological relic density has to be in 
the the 99\% confidence level WMAP narrow range, that is 
\begin{equation}
0.087 \leq \Omega_{\chi}\, h^2 \leq 0.138
\label{omrange}
\end{equation}
where $\Omega \equiv \rho / \rho_c$ with $\rho_c \simeq 2 \cdot 10^{-29} h^2
{\rm g/cm^3}$ is the ``critical'' mass density that yields a flat universe and 
the dimensionless parameter $h$  the scaled Hubble constant describing  the
expansion of the universe. The result for the relic density  has been obtained
using the code {\tt microMEGAS} \cite{micromegas}. \bigskip

We also take into account the most recent astrophysical bounds, in addition  to
the WMAP constraint. For the evaluation of the neutralino--nucleon cross section
for direct detection and the gamma--ray fluxes relevant for the indirect
detection, we use the latest released version of the program {\tt DarkSUSY}
\cite{darksusynew}. We have included the relevant sensitivities and
uncertainties in both  detection methods.  The measurement of the  WIMP--nucleon
cross section depends on the speed of the WIMP candidate in the neighborhood of
the sun and  it has been shown in Ref.~\cite{Bottino:2005qj} that the
astrophysical uncertainties can affect the experimental sensitivities by a
factor of two. Moreover, the modelization of the  $\pi$--nucleon can induce
theoretical uncertainties in the  calculation of the scattering WIMP--nucleon
cross section, estimated to be a factor five at its maximum \cite{Ellis:2005mb};
rather large QCD corrections and uncertainties are also present \cite{QCD}.   We
will assume in this study an uncertainty  factor of 2 coming from the
astrophysical uncertainty on the WIMPs speed in our galaxy and a factor of 3 in
the calculation of $\sigma (\neut N \to \neut N)$. Concerning LSP indirect
detection, most of the uncertainties are coming from the  model used for the
dark matter halo.  For illustration, we will use in this paper a cuspy Navarro
Frenk and White (NFW) profile \cite{Navarro:1996he}, keeping in mind that a
cuspier one [such as the one proposed by Moore in Ref.~\cite{Moore:1999gc}] or
a smoother (Isothermal) one, will give fluxes that are two orders of magnitude
higher and lower, respectively.

 \subsubsection*{3.2 The spectrum in the intense coupling regime}

In the present study, we choose for illustration two NUHM scenarios. In the 
first one,  we fix $\tan\beta=30$, the common
sfermion mass to $m_0=2$ TeV, the common gaugino mass to $M_{1/2}=180$ GeV and
the trilinear coupling in such a way that $A_t= 3$ TeV to maximize the mixing
in the Higgs sector and thus the mass of the lightest $h$ boson in order to
pass easily the LEP2 constraint $M_h \gsim 114$ GeV in the decoupling regime
when the theoretical uncertainty of $\sim 3$ GeV is included. We also fix the
non--universality parameter to $\delta_1=-1$, but vary the other parameter
$\delta_2$ in a narrow range to have a pseudoscalar $A$ boson mass of the
same order as the masses of the CP--even $h$ and $H$ bosons. In a second
scenario, we choose a slightly lower value of $\tan\beta$, $\tan\beta=20$, and
a larger value of the universal gaugino mass, $M_{1/2} =580$ GeV; all the other
parameters are the same as in the previous case. \smallskip

\begin{figure}[h!]
\vspace*{-.4cm}
    \begin{center}
	\hskip -.3cm
       \epsfig{file=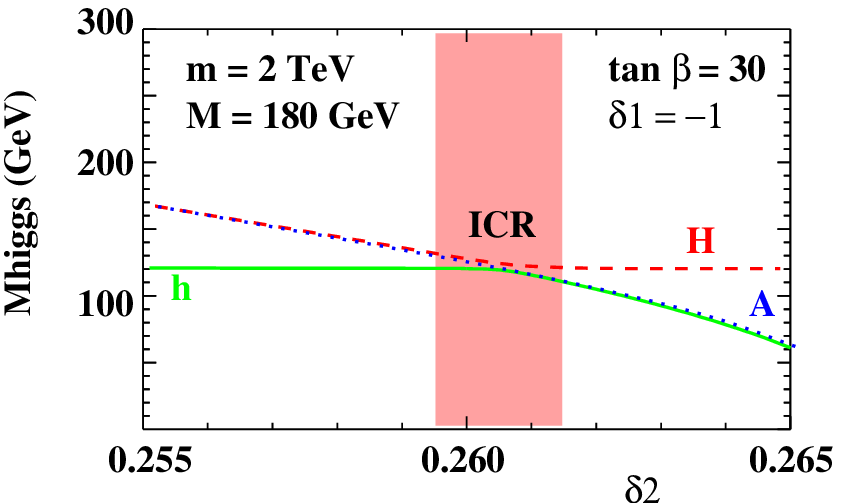,width=0.5\textwidth}
	\hskip .1cm
       \epsfig{file=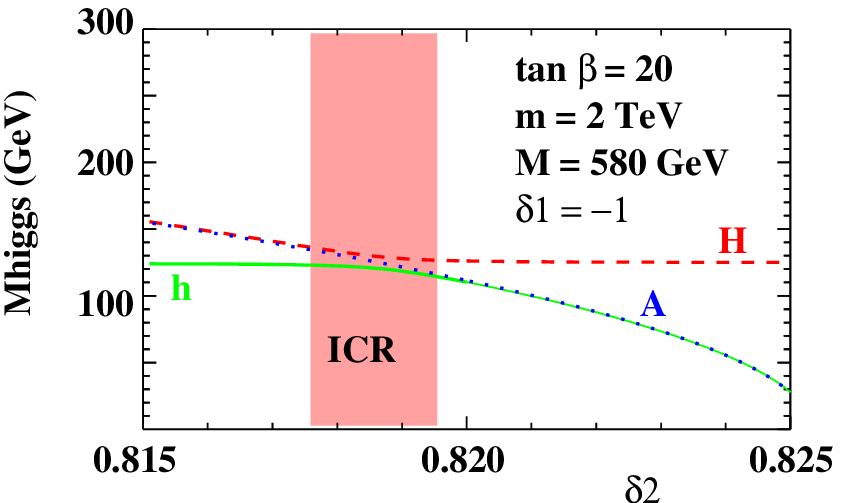,width=0.5\textwidth}
       \vskip .4cm
       \epsfig{file=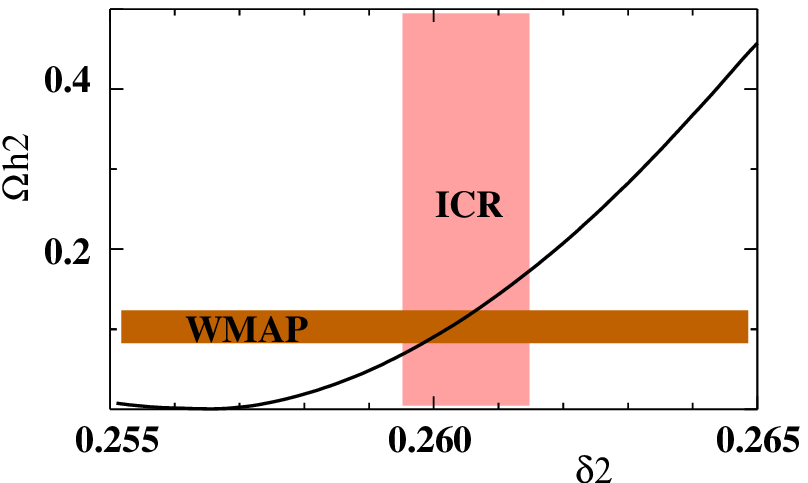,width=0.50\textwidth}
	\hskip -.3cm
       \epsfig{file=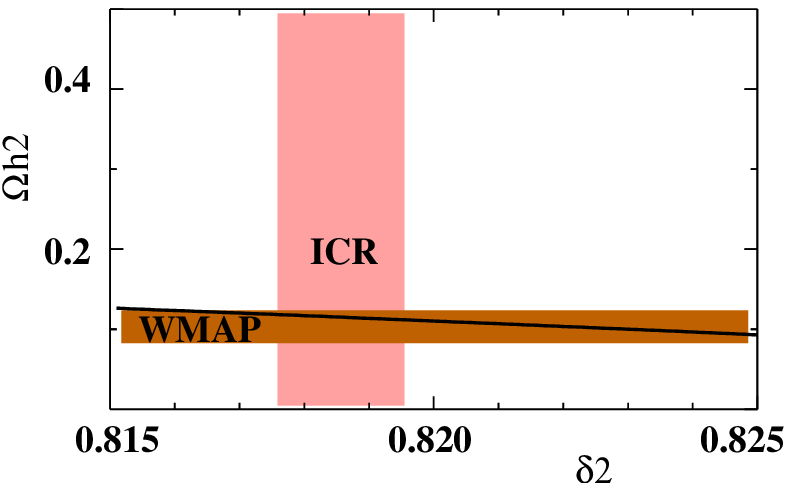,width=0.5\textwidth}
	\vskip -0.1cm
\caption{The neutral Higgs boson masses as a function of the parameter 
$\delta_2$ in the two scenarios discussed in the text (top) and the
cosmological relic density of the LSP neutralino (bottom). The pink vertical 
bands display the intense coupling regime while the brown horizontal lines 
show the WMAP range for the LSP relic density.}
        \label{fig:masses}
    \end{center}
\vspace*{-.5cm}
\end{figure}

In the upper part of Fig.~2, we display the masses of the three neutral Higgs
bosons as a function of the parameter $\delta_2$ in the two scenarios. 
The pink vertical bands correspond to the areas in which $|M_{H,h} -M_A| \leq 
10$ GeV, that is, to the intense coupling regime which appears around $\delta_2
=0.260$ in the first case and $\delta_2=0.818$ in the second case.  These 
two values of  $\delta_2$ lead to an almost mass degeneracy of the neutral 
Higgs particles $M_h\sim M_H \sim M_A \sim 120$ GeV and define the points
{\bf A} and {\bf B} corresponding to the initial values at the GUT scale :
$m_{H_1}=0$ and, respectively, $m_{H_2}=1.26 m_0$ and $m_{H_2}=1.818 m_0$. In
the lower part of Fig.~2, we display the cosmological relic density of the LSP
neutralino, again as a function of $\delta_2$, and compare it to the WMAP
narrow band. For the values of $\delta_2$ which correspond to the points  {\bf
A} and {\bf B}, we see that the WMAP constraint is obeyed.\smallskip 

In the two scenarios, the output for the Higgs boson masses are shown in 
Table~1 for the values of the parameters $\delta_2$ which correspond to the
selected points {\bf A} and {\bf B}. We also display the  masses of the two
lighter neutralinos $m_{\chi_1^0}$ and $m_{\chi_2^0}$ and the value of the
parameter $\mu$. As can be seen, in point {\bf A}, the LSP $\chi_1^0$  is rather
light while the $\mu$ parameter is very large: the LSP is thus almost bino--like
and the lighter  chargino is almost mass  degenerate with the wino--like $
\chi_2^0$, while the heavier neutralinos and chargino are higgsino--like and
have masses close to $\mu$. In point {\bf B}, the values of the gaugino  and 
higgsino mass parameters are of the same order, $M_2 \approx \mu$, and
all neutralinos and charginos are thus mixed states with masses that are
quite close to each other. The LSP neutralino has a mass which is relatively 
large, exceeding the $W/Z$ boson and all Higgs  boson masses and even the top 
quark mass. \smallskip

In Table 1, we also display the result for the LSP relic density $\Omega_\chi
h^2$ and for the fractions of the various final states which contribute,
B$(\neut \neut \rightarrow X) \equiv \sigma (\neut \neut \rightarrow X)/
\sigma( \neut \neut \rightarrow {\rm all})$. As we have assumed heavy scalar
fermions in both scenarios, $m_0=2$ TeV, the main annihilation channel for the
LSP neutralinos, leading to a relic density compatible with the WMAP result, is
the Higgs boson [mainly $A$ and to a lesser extent $h/H$] exchange
contributions in the case  of point {\bf A}. As can be seen in the table, in
this case, the pseudoscalar $A$ boson mass is close to twice the LSP neutralino
mass and the $A b\bar b$ coupling is very strong for $\tan\beta=30$; this
enhances the annihilation process  $\chi_1^0 \chi_1^0 \xrightarrow{A} b\bar{b}$
which reaches a branching fraction of 90 \%. For point {\bf B} with larger
values of $M_{1/2}$ and $\mu$, higher values of the parameter $\delta_2$ are
needed to increase $m_{H_2}$ and keep a light pseudoscalar $A$ boson, as can be
seen from eq.~(\ref{ma2}). Since here, the $A$--pole is too far from  the $2
m_{\chi_1^0}$ threshold, the contribution of the annihilation process $
\chi_1^0 \chi_1^0 \rightarrow b\bar{b}$ to the LSP relic density is only at 
the level of $10\%$; the $ \chi_1^0 \chi_1^0    \to t\bar{t}$ annihilation
process, which has roughly the same probability as $ \chi_1^0 \chi_1^0    \to
b\bar{b}$, occurs mainly through $Z$ boson exchange.  In fact, since in this
scenario one has $m_{\chi_1^0} \sim 250$ GeV, all Higgs particles can occur as
final states in the annihilation of the LSPs and, since the LSP is a
bino--higgsino mixed state with strong couplings to the Higgs bosons, the
fraction of the  processes involving at least one final Higgs particle is
larger than 70\%. \smallskip

\begin{table}[!h] 
\vspace*{-.1cm}
\begin{center} 
\renewcommand{\arraystretch}{1.25}
\begin{tabular}{|c|c|c|c||c|c|} \hline  
&inputs& $M_{\rm \Phi}$ & $m_{\chi_i}$ & $\Omega_\chi h^2$ & BR($\neut \neut 
\rightarrow X$) \\  \hline  
& $\tan \beta= 30$ & $M_h=120.4$ GeV & $m_{\chi_1^0}= 78$ GeV & & 
$b \bar b : 90\% $ \\
{\bf A}& $\delta_2= 0.260$ & $M_H=126.5$ GeV & $m_{\chi_2^0}= 155$ GeV& 0.09&
$\tau\tau : 10\%$ \\
& $M_{1/2}=180 $ GeV & $M_A=126.4$ GeV &$\mu=846$ GeV & & \\ \hline
&&&&& $b\bar b (t \bar t) : 13\% ( 12 \%) $ \\
& $\tan \beta= 20$ & $M_h=124$ GeV & $m_{\chi_1^0}= 241$ GeV & & $hA (hh): 
27\% (1\%) $\\
{\bf B} & $\delta_2= 0.818$ & $M_H=136$ GeV & $m_{\chi_2^0}= 351$ GeV &
0.12 & $ZH (Zh) : 16\% (5\%)$ \\
 & $M_{1/2}=580 $ GeV & $M_A=134$ GeV &$\mu=362$ GeV & & $W^\pm H^\mp : 22\%$ 
\\  &&&&& $WW : 2\% $ \\
 \hline
\end{tabular} 
\caption{Sample spectra obtained in the non--universal scenarios {\bf A} and 
{\bf B} (with some of the parameters indicated and $m_0=2$ TeV), together with 
the value of the neutralino relic density and the fractions of the main 
subprocesses contributing to it. } 
\label{tab1}
\end{center}
\vspace*{-.5cm}
\end{table} 

Finally, in Fig.~3, we show the result of a scan in the ($M_{1/2}$, $\delta_2$)
plane,  starting with the basic inputs for $m_0=2$ TeV, $\tb=30$ (left) and 20
(right) and $\delta_1=-1$; the position of the points {\bf A} and {\bf B}
analyzed earlier are marked by a cross. In the white area, all the points
fulfill the experimental and theoretical constraints discussed previously,
except for the WMAP constraint which is  obeyed only in the red narrow strips
[between the strips, the relic density is either too high or too low, compared
to the WMAP value].  The intense coupling regime, defined by the requirement
$|M_{H,h}-M_A| <10$ GeV, is represented by the thin black lines. In the (green)
areas above these lines, the lightest $h$ boson has a mass smaller than the
LEP2 limit, $M_h \lsim 114$ GeV.  One notices that increasing the value
of $\delta_2$ for a given value of $M_{1/2}$, finally leads to a region in
which $M_A^2 < 0$ : the $m_{H_2}^2$ term is not sufficiently negative at the
electroweak scale to compensate the $m_{H_1}^2$ term in eq.~(\ref{ma2}).

\begin{figure}[!h]
    \begin{center}
       \epsfig{file=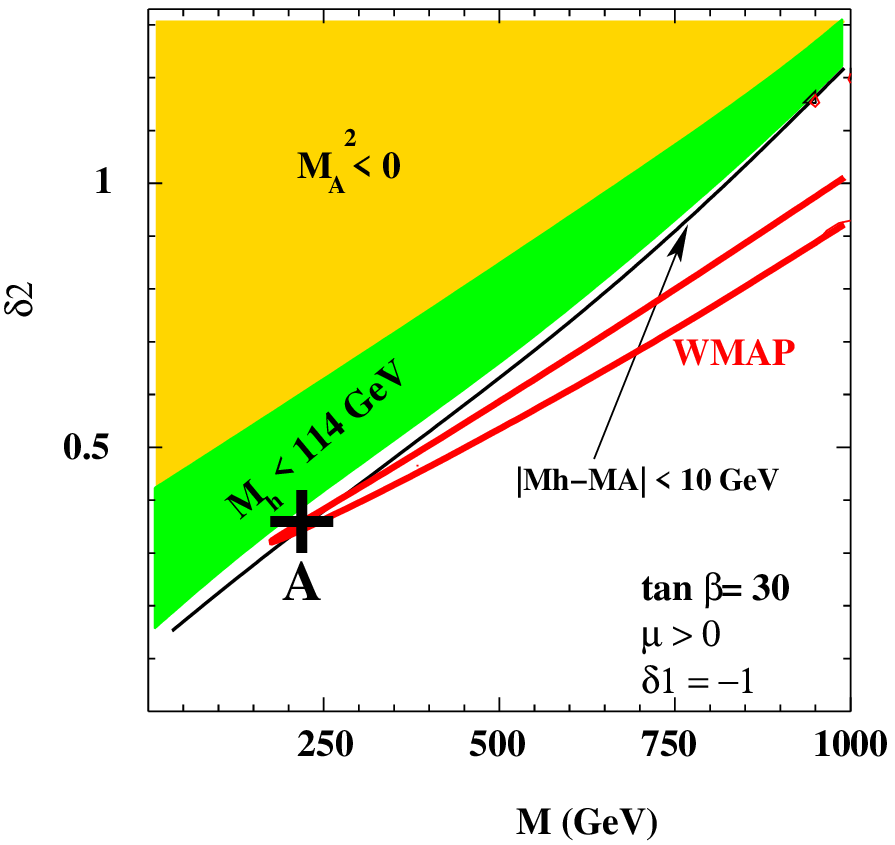,width=0.5\textwidth}
	\hskip -.3cm
       \epsfig{file=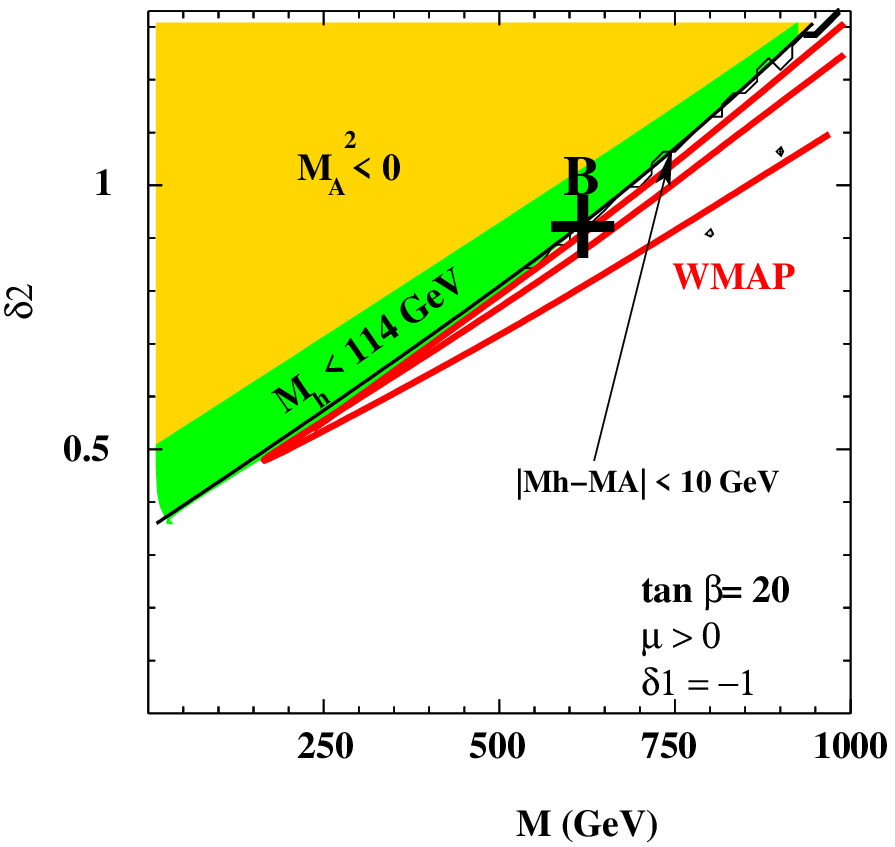,width=0.5\textwidth}
	\vskip -0.1cm
\caption{The allowed ($M_{1/2}$, $\delta_2$) parameter space in the NUHM model 
with $\tan \beta=30$ (left) and $\tan \beta=20$ (right), $\mu > 0$, $A_t=3$ 
TeV, $m_0=2$ TeV and $\delta_1=-1$. We indicate in the plot the points {\bf A} 
and {\bf B} defined in the text, the WMAP range, the area in which $M_h \leq
114$ GeV, and the intense coupling regime (black solid lines).}
        \label{Scan}
    \end{center}
    \vspace*{-.5cm}
\end{figure}

\subsubsection*{3.3 Prospects for dark matter detection}

Let us now turn to the prospects for detecting the LSP dark matter candidate in
these two scenarios. In direct detection, as a consequence of the light Higgs
bosons with strongly enhanced couplings to down--type quarks that are present,
the scalar or spin--independent  LSP--nucleon  scattering cross section can be
extremely large as the process is mainly mediated by the exchange of the two
CP--even Higgs bosons $h$ and $H$. As can be seen in the upper part of Fig.~4,
where the expected $\neut N \to \neut N$ scattering cross section and the
CDMS--Soudan sensitivity in which the various uncertainties discussed in
section 2.3 are incorporated, the intense--coupling regime is within the reach
of the CDMS--Soudan detector. For instance, in point {\bf A} with
$\delta_2=0.260$, one has $\sigma (\chi_1^0 N \to \neut N)= 1.2 \times 10^{-7}$
pb which is only slightly lower than the present CDMS sensitivity of  $4\times
10^{-7}$ pb and much larger than the expected sensitivity of $2\times 10^{-8}$
pb. In the second scenario, one obtains an even larger cross section despite of
the larger value of the LSP mass and the smaller value of $\tan\beta$. For
instance, in point {\bf B} with  $\delta_2=0.818$, one obtains $\sigma
(\chi_1^0 N)= 10^{-6}$ pb. This is due to the fact that here, the LSP is a
mixed bino--higgsino state with larger couplings to the Higgs bosons [in the
first scenario, the larger value of $\tan\beta$ and the smaller LSP mass were
partly compensated by the smaller Higgs couplings to the almost bino--like
neutralino].  Note that such high values for the LSP--nucleon scattering cross
section have also been obtained in Ref.~\cite{Nonunivdirect} in the case of
light pseudoscalar Higgs bosons not pertaining to the intense--coupling 
regime.\smallskip

\begin{figure}[!h]
\vspace*{-.5cm}
    \begin{center}
	\hskip -.1cm
       \epsfig{file=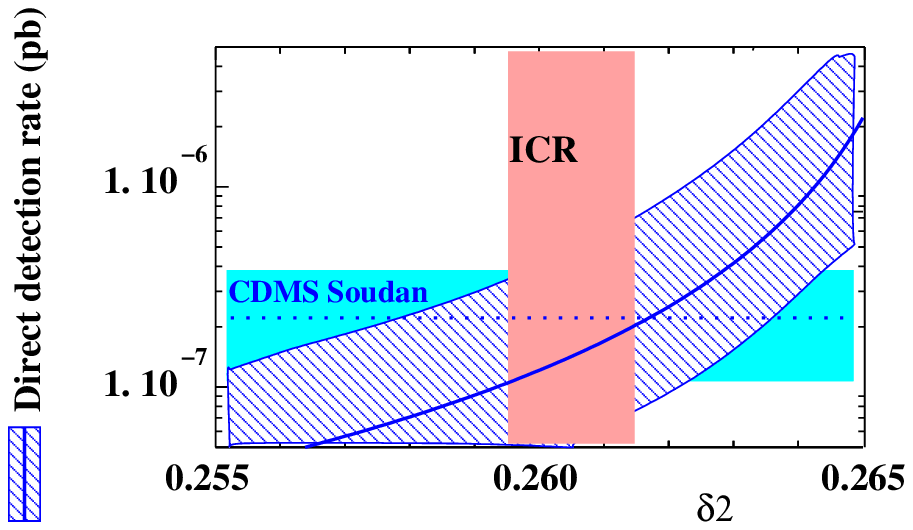,width=0.49\textwidth}
	\hskip .2cm
       \epsfig{file=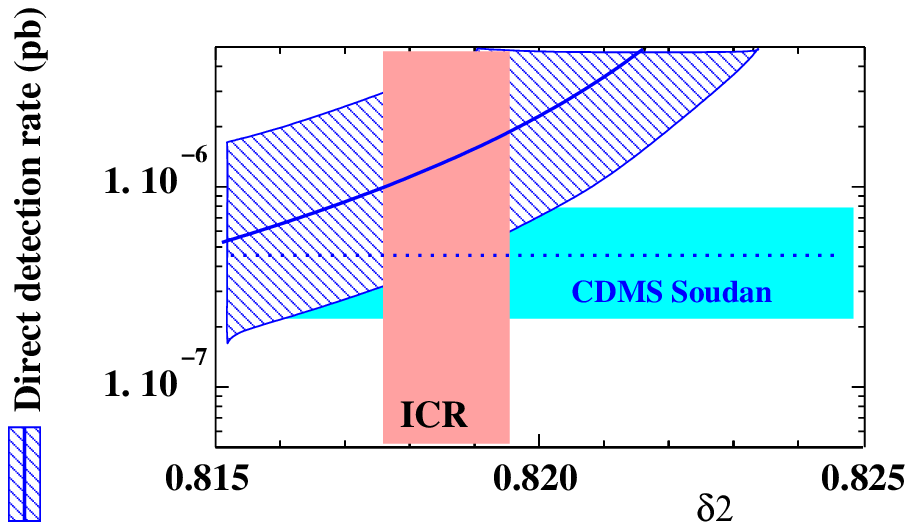,width=0.49\textwidth}
	\vskip 0.5cm
	\hskip -.3cm
      \epsfig{file=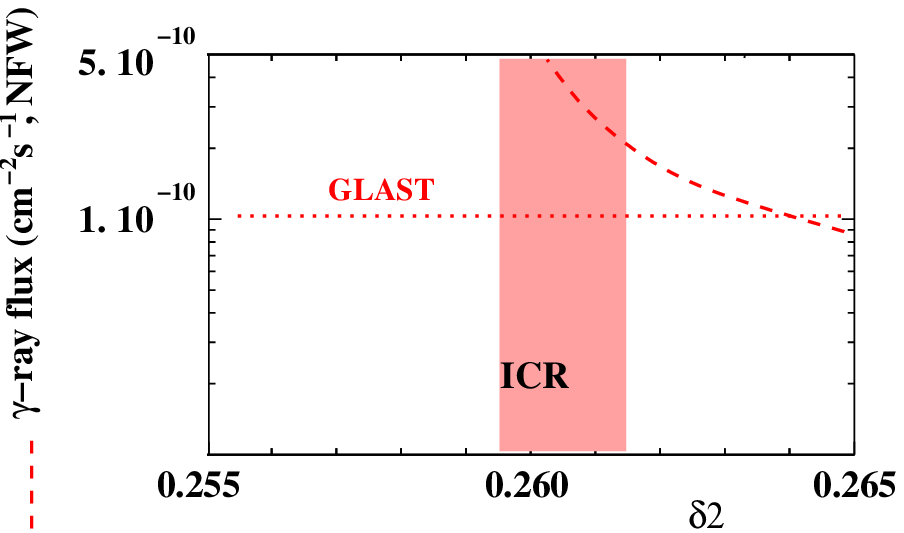,width=0.49\textwidth}
	\hskip .2cm
       \epsfig{file=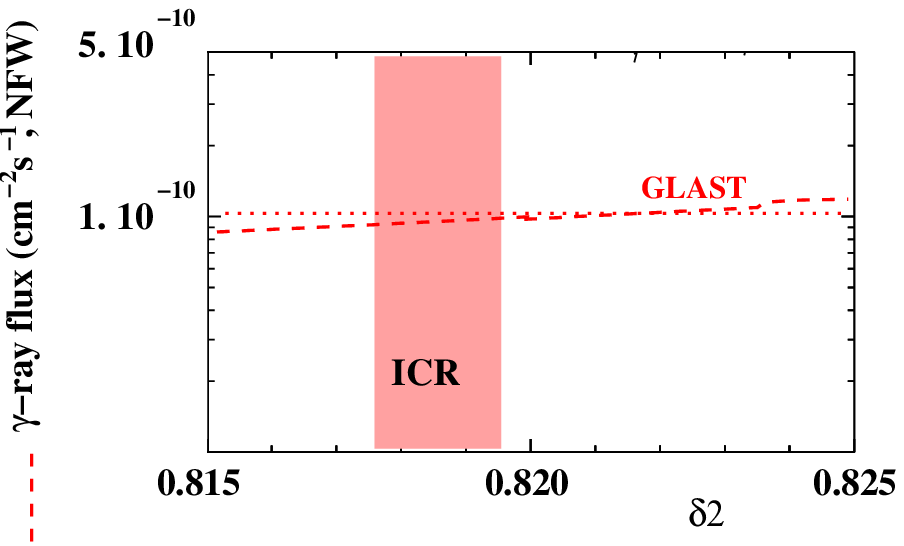,width=0.49\textwidth}
\caption{Upper curves: the rate for the LSP--nucleon cross section including
the various errors (hatched bands) compared to the expected sensitivity of the
CDMS--Soudan experiment (horizontal blue bands) as a function of the
non--universality parameter $\delta_2$. Lower curves: the obtained  gamma ray
flux in photons $\mathrm{cm^{-2} s^{-1}}$ using the NFW profile  as a function 
of $\delta_2$ (dashed curves) and the expected sensitivity of the GLAST 
experiment (dotted curves). The left (right) handed plots are for the scenarios
discussed previously with $M_{1/2}=180$ GeV (580 GeV) and the other parameters 
as described in the text.}
\label{fig:detection}
    \end{center}
\end{figure}

Concerning indirect LSP detection, we have calculated the rate in the framework
of the NUHM model, assuming the NFW profile \cite{Navarro:1996he} within a 
solid angle of $\Delta \Omega= 10^{-5}$ sr and found in the two scenarios 
fluxes around or larger than $\sim 10^{-10} \mathrm{cm^{-2} s^{-1}}$, which is 
within the reach of a detector like GLAST. This is shown in the lower parts of 
Fig.~4, where we display the obtained  gamma ray flux as a function of 
$\delta_2$ and compare it to the expected sensitivity of the experiment. To be 
more precise, we find a flux of $\Phi_{\gamma}^{\rm NFW}= 5.6 \times 10^{-10}
\mathrm{cm^{-2} s^{-1}}$ for point {\bf A} and $\Phi_{\gamma}^{\rm NFW}= 8.4 
\times 10^{-11} \mathrm{cm^{-2} s^{-1}}$, for point {\bf B}. In the first case,
the obtained flux is much larger than in the second case because of the stronger
enhancement of the annihilation channel through $A$--boson exchange and  the
smaller value of the LSP neutralino mass. \smallskip

Thus, one can conclude that the two scenarios that we have chosen to illustrate
the Higgs intense coupling regime in the NUHM model are in a sense
orthogonal, as point {\bf A} will be more easily probed in indirect detection
experiments whereas point {\bf B} can be best probed in direct LSP detection.

\subsection*{4 The mSUGRA case}

For completeness, let us briefly discuss the intense coupling regime in the
mSUGRA framework with universal Higgs mass parameters. As mentioned earlier,
the pseudoscalar Higgs boson mass and the $\mu$ parameter tend to be quite
large in this case, except in the focus point region when both the common
scalar mass $m_0$ and $\tb$ are large. This, however,  occurs only in a tiny 
part of the parameter space near the region where electroweak symmetry breaking
is not achieved [in general because either $M_A^2<0$ or $\mu^2 <0$ or both].  
The Higgs intense coupling regime occurs thus in an even smaller region of the
 parameter space and one has to tune severely some of the input parameters to 
realize it.\s

For some of the input parameters already used in the NUHM model, $m_0=2$ TeV
and $A_t=3$ TeV, and assuming $\mu>0$ and $M_{1/2}= 350$ GeV for instance, one
obtains using the program {\tt SuSpect}, the intense coupling regime only for
values of $\tb$ very close to $\tb= 53.87$ as is exemplified in Fig.~5, where
the neutral Higgs boson masses are shown in this mSUGRA scenario as a function
of $\tb$. For the particular value $\tb =53.87$, one has a pseudoscalar Higgs
boson mass of $M_A \simeq 122$ GeV leading to $M_h \simeq 117$ GeV and
$M_H\simeq 124$ GeV for the masses of the CP--even Higgs particles. For the
chosen inputs, one also obtains a relatively large value for the higgsino mass
parameter, $\mu \simeq 440$ GeV, with a lightest neutralino mass of $m_{\neut}
\simeq 150$ GeV. The LSP is thus almost bino--like with only a small higgsino
component. However, this component is large enough to lead to a substantial
annihilation rate of the LSP neutralino into Higgs bosons. Indeed, one obtains
for instance B$(\neut \neut \to hA+HA$) and B$(\neut \neut \to HZ+hZ)$ of,
respectively, 12\% and 8\% the dominant channel being annihilation into $b\bar
b$ and $\tau^+ \tau^-$ final states which has an overall fraction close to
70\%. The obtained value for the LSP cosmological relic density in this
scenario is $\Omega_\chi h^2 \simeq 0.1$ which is within the WMAP range. 
Because of the very high value of $\tb$, the couplings of the rather light
CP--even neutral Higgs bosons $h$ and $H$ to $b\bar b$ states are extremely
strong, leading to a large LSP--nucleon scattering cross section, $\sigma
(\chi_1^0 N \to \neut N) \sim 2 \times 10^{-6}$ pb, which can be easily probed
by the CDMS experiment. The indirect detection rate of the lightest neutralino
is also significant as one obtains a gamma--ray flux of the order of
$\Phi_{\gamma}^{\rm NFW} \sim 2 \times 10^{-10} \mathrm{cm^{-2} s^{-1}}$ which
can be observed by GLAST.\s

\begin{figure}[!h]
    \begin{center}
\hskip -.2cm
\mbox{\epsfig{file=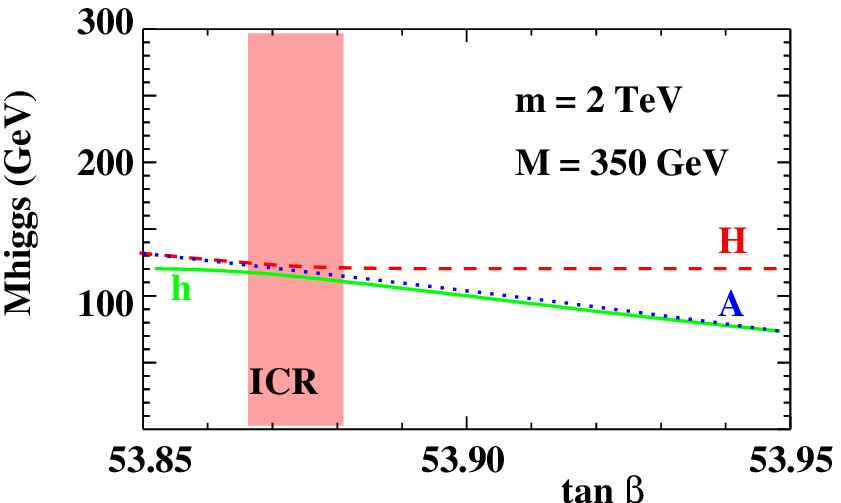,width=0.48\textwidth}
      \hskip .2cm
      \epsfig{file=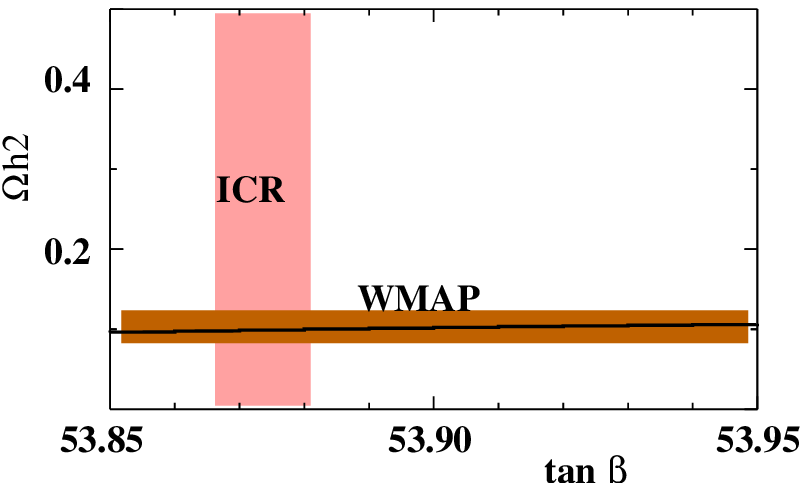,width=0.48\textwidth} }
	\vskip 0.3cm
	\hskip -.2cm
\mbox{\epsfig{file=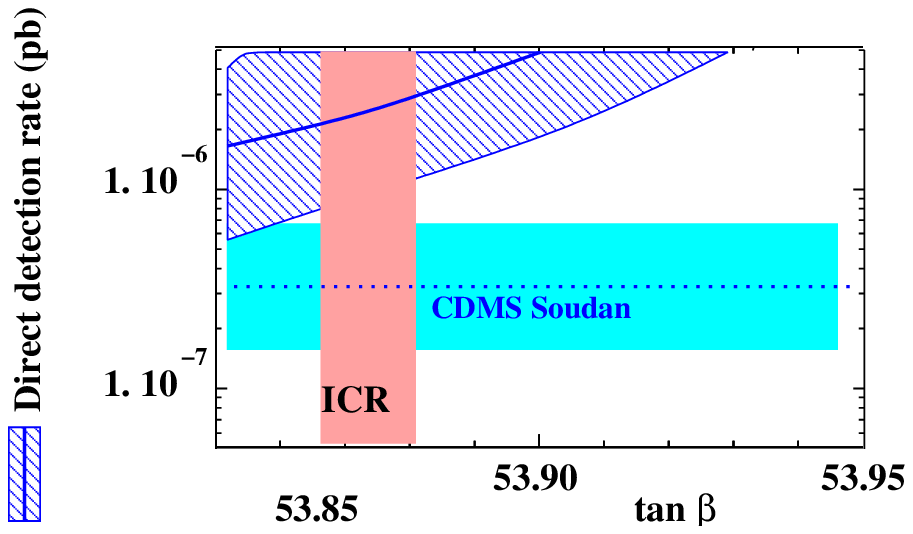,width=0.48\textwidth}
      \hskip .2cm
       \epsfig{file=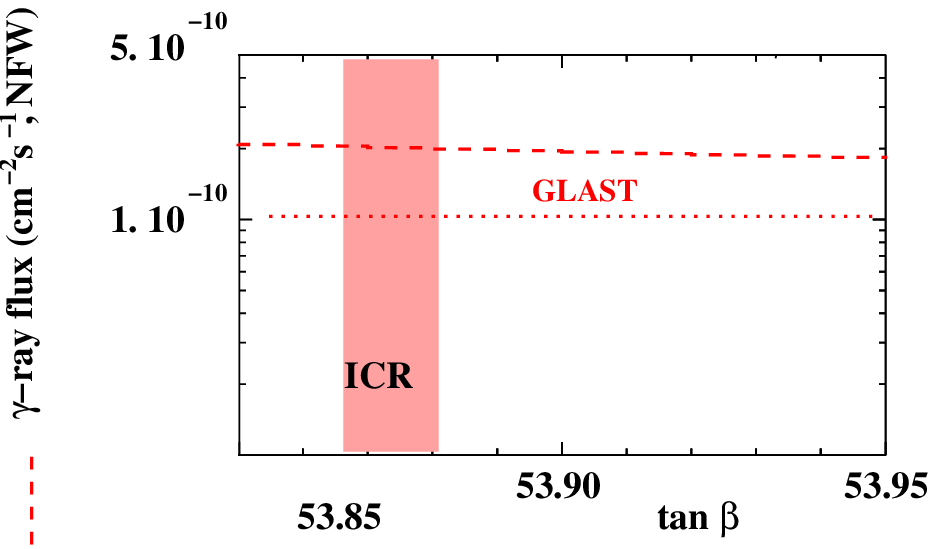,width=0.48\textwidth} }
          \caption{The same as in Figs.~2 and 4 but for the mSUGRA  model
with universal Higgs mass parameters and as a function of $\tb$ with basic 
inputs $m_0=2$ TeV, $A_t=3$ TeV, $M_{1/2}=350$ GeV and $\mu>0$.}
        \label{fig:mSUGRA}
    \end{center}
\vspace*{-0.3cm}
\end{figure}

However, there is a potential problem with this particular scenario.  For such
a large value of $\tb$ and a low pseudoscalar Higgs boson mass $M_A$, the decay
rate B($B_s \to \mu^+ \mu^-) \propto \tan^6\beta/ M_A^2$ is huge [as already
noticed in Ref.~\cite{Baek:2005wi} for instance] and one obtains for the point
introduced above B($B_s \to \mu^+ \mu^-) \sim 4 \times 10^{-5}$ to be compared
with the Tevatron experimental bound of B($B_s \to \mu^+ \mu^-) \sim 2.9 \times
10^{-7}$ \cite{bmumuexp}. As discussed in section 3.1, the two--orders of
magnitude discrepancy between these two values will need a rather strong flavor
mixing to be resolved. [Note that we have a similar problem for the point {\bf
A} of the NUHM model, but the difference between the obtained $B_s \to \mu^+
\mu^-$ rate and the experimental bound is much smaller and can be attributed to
theoretical uncertainties.]\s

Nevertheless, if this problem can be circumvented, one would be very close to
definitely test this scenario. Indeed, the cross section for the associated
production of the neutral Higgs bosons in association with $b\bar b$ pairs,
$p\bar p \to gg/q \bar q \to b\bar b + A,h,H$, will be extremely large at the
Tevatron for the considered values of $\tb$ and $M_A$.  With the luminosity
that is presently collected at the Tevatron, values $\tan \beta \sim 50$
are already ruled out for $M_A \sim 100$ GeV \cite{Tevatron} and the 2007
projection for the increase of the Tevatron luminosity would allow to observe
or rule out a Higgs boson signal in this process for pseudoscalar Higgs bosons
masses of $M_A \sim 120$ GeV \cite{Carena:2006dg}.

\subsection*{5. Conclusions}

In this paper, we have investigated the possibility of realizing the intense
coupling regime, in which  all the Higgs bosons of the MSSM have comparable
masses and the parameter $\tb$ is large, in the framework of the constrained
MSSM.  We have pointed out that this interesting regime is possible within the
mSUGRA model, but when all the scalar fermion and Higgs soft SUSY--breaking
mass terms are unified at the GUT scale, this happens in a very limited range
of the parameter space: large values of $\tb$, $\gsim 50$, are needed which
might lead to a severe conflict with the measured value of the $B_s \to \mu^+
\mu^-$ branching rate and potentially, with the search of the MSSM Higgs bosons
at the Tevatron. In contrast, relaxing the universality of the SUSY--breaking
scalar masses of the two MSSM Higgs fields at the high scale, $m_{H_1} \neq
m_{H_2} \neq m_0$, leads to more freedom which allows to reach this regime in a
much larger area of the mSUGRA parameter space. We have provided examples of
scenarios in which the intense coupling regime is realized in the NUHM model
and where, not only all known theoretical requirements and collider experiment
constraints are fulfilled, but also where the generated relic density of the
lightest neutralino of the MSSM is compatible with the WMAP measurement of the
amount of cold dark matter in the universe. \s 

 These scenarios have a very interesting phenomenology. They lead by definition
to light MSSM Higgs bosons which can be observed at the next generation of
colliders such as the LHC and the ILC [and also at the Tevatron if the value of
$\tb$ is very large]. In addition, because of the large $\tb$ and low $M_A$
values which are implied by these models, the rates for direct and indirect
detection of the LSP cold dark matter neutralino are large enough for this
particle to be observed in near future experiments such as CDMS--Soudan and
GLAST.  Furthermore, since at least the spectrum of spin--1/2 superparticles
needed to generate the right cosmological density for the LSP is relatively
light, the heavier charginos and neutralinos [and possibly the gluinos] could
also be produced at the next generation of collider experiments. This would
allow to precisely measure the couplings of these particles, in particular the
couplings to the Higgs bosons in possible cascades \cite{cascades}, and determine the
parts of the MSSM Lagrangian which enter in the expressions of the relic
density and the detection rates. Thus, using the future collider data, one
would be able to check that the LSP neutralino makes indeed the entire dark
matter of the universe and that the detection rates observed in astrophysical
experiments are indeed those predicted by the model.  In this respect, these
scenarios, if realized in Nature, would highlight the complementarity between
collider physics searches and precision measurements and astroparticle physics
searches.  

\vspace*{1cm}

\noindent{\bf Acknowledgments:}\smallskip 

\noindent This work is supported by an ANR (Agence 
Nationale pour la Recherche) grant for the project {\tt PHYS@COL\&COS} under 
the number NT05-1\_43598. The work of Y.M. is sponsored by the PAI 
program PICASSO under contract PAI--10825VF. We thank Emmanuel Nezri for 
discussions.

\nocite{}
\bibliography{bmn}
\bibliographystyle{unsrt}

\end{document}